# Crosstalk in multi-output CCDs for LSST

**P. O'Connor**

*Brookhaven National Laboratory,*
*Upton, NY USA*
*E-mail*: poc@bnl.gov

ABSTRACT: LSST's compact, low-power focal plane will be subject to electronic crosstalk with some unique signatures due to its readout geometry. This note describes the crosstalk mechanisms, ongoing characterization of prototypes, and implications for the observing cadence.

## Contents

## 1. The LSST camera and front-end electronics

LSST's focal plane readout is highly parallelized to achieve 3.2Gpixel readout in 2s [1,2]. Each of the 189 4Kx4K CCDs in the science array has 16 segments with independent output amplifiers; a total of 3024 video channels are read out synchronously using custom CMOS ASICs. As illustrated in Figure 1, the entire readout chain is located inside the cryostat vacuum to allow the shortest possible connection length between the CCDs and front end electronics and to limit the camera's obscuration of the telescope beam. This arrangement constrains the volume to around 50l and the power dissipation budget to around 1kW. The table in Figure 1 shows the resulting performance requirements on the readout electronics. The combination of high speed, high-resistivity silicon, low power, and tight channel spacing make the LSST readout more susceptible to electronic crosstalk than previous mosaic cameras.

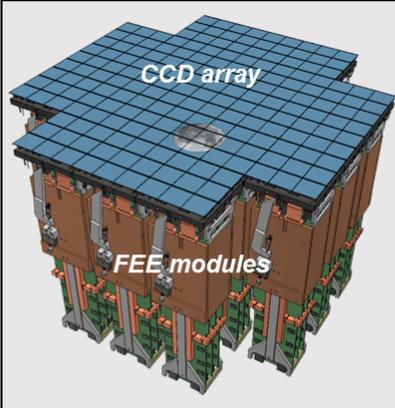

**Figure 1: LSST focal plane parameters.**

## 2. Crosstalk mechanisms

The major contributor to crosstalk is capacitive coupling between single-ended video outputs from the CCD. Each CCD output couples to neighboring channels via capacitance between amplifiers on the silicon CCD and between traces on the CCD package, flex cables and front-end PCB. A simple model for the electrical network between an aggressor and victim channel is shown in Figure 2. In this model the aggressor ($V_1$) pulse couples to the victim ($V_2$) via the coupling capacitance $C_C$. The victim amplifier (without signal) is represented by its output impedance $R$ (the same as the aggressor's output impedance). The victim amplifier is loaded by a capacitance $C_L$. In the frequency domain the crosstalk $V_2/V_1$ is given by

$$\frac{V_2}{V_1} = \frac{sC_C R}{1 + s(C_C + C_L)R}$$

where $s = j\omega$ is the frequency. The magnitude of the crosstalk increases with higher frequency, larger coupling capacitance, or higher output impedance. These factors are exacerbated by LSST's fast readout, small channel separation, and low power respectively.

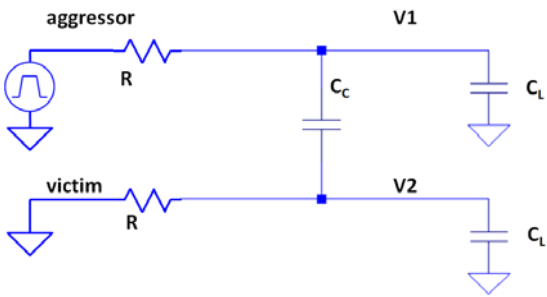

**Figure 2: aggressor-victim coupling.**

In more detail, the correlated double sampling circuit of the front end electronics needs to be considered. This is shown in Figure 3 along with a time domain simulation. The CDS circuit in LSST uses dual slope integration (DSI), which takes the difference of the integrated signal at the CCD output before and after the arrival of charge on the sense node. In the time domain, the large step-like signal on the aggressor capacitively couples to the victim, producing a pulse-like signal with an exponential decay. The second DSI integration starts with a delay $T_d$ following the aggressor step. If the second integration starts before the victim signal has fully decayed, a remnant of the decaying tail of the victim's crosstalk response is integrated. The crosstalk amplitude after CDS thus depends on the DSI delay time $T_d$ relative to the decay time $RC_L$ of the victim channel.

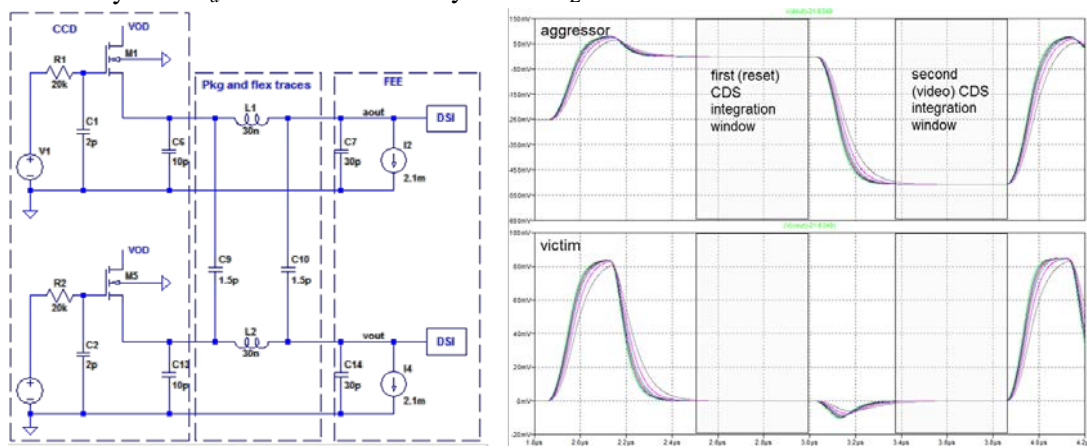

**Figure 3: Detailed coupling circuit (left); aggressor and victim waveforms (right).**

We also investigated the contribution of non-capacitive coupling. The mechanisms considered included (1) coupling via parasitic impedance in the common output drain bias circuit, and (2) common impedance in the first-stage followers' source-to-front substrate ground. Mechanism (1) was ruled out because the expected pattern of crosstalk would be different from the one observed. To evaluate mechanism (2), we measured the resistance between separate contacts to the CCD ground (finding no resistance greater than 5 Ohms) and estimated resistance of the tracks on the multilayer ceramic carrier (again finding R>10Ohms implausible). These parasitic impedances were added to the SPICE model, which showed that the magnitude of crosstalk found experimentally could only be explained if those resistances were an order of magnitude larger.

## 3. Characterization of sensor prototypes

### 3.1 Requirements

During survey operations, bright sources being imaged on one section of a CCD will produce crosstalk artifacts in other sections. These artifacts ("ghosts") can be flagged since their location relative to the aggressor will always be known. If the aggressor-to-victim coupling coefficients are known then the crosstalk can be corrected in software.

Let's make a rough estimate of the number of ghosts in a typical LSST image. LSST's single-visit $5\sigma$ limiting magnitude is $r = 24.2$; saturation occurs at $r \sim 16$ [3]. If the crosstalk level is $x$ then an $r < (24.2+2.5\log(x))$ source will produce a detectable crosstalk ghost. Because the full well capacity of the LSST sensors is limited, sources brighter than $r = 16$ will saturate and start to bloom rather than produce larger aggressor signals. This means that no source can produce a detectable ghost if the crosstalk level is less than 0.05% (8.2 mag). Of course, ghosts which are below the detection limit are still a concern as they may compromise the image in other ways; and the ghosts of bright, blooming aggressor sources will contaminate larger sections of the CCD than non-bloomed sources.

To illustrate the situation, Figure 4 shows the expected number of bright stars capable of producing detectable crosstalk ghosts as a function of crosstalk amplitude. Estimates of star counts vs. magnitude were taken from [4] at mid galactic latitudes. The discontinuity at 0.05% is due to the saturation limit. Also shown in the figure are LSST's requirement and goal for crosstalk between amplifier segments.

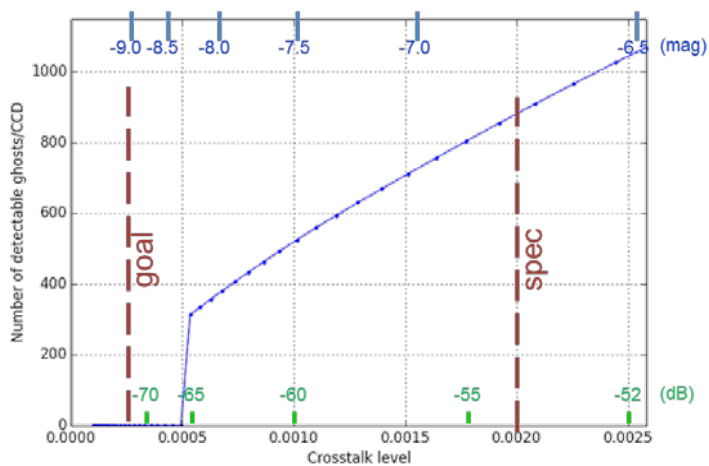

**Figure 4: Number of sources capable of producing detectable crosstalk ghosts per CCD, as a function of crosstalk amplitude. LSST goal and specification indicated by vertical dashed lines.**

Crosstalk can occur between amplifier segments within a single CCD and between CCDs in the focal plane. In LSST's focal plane the CCDs and FEE are organized into rafts of 9 CCDs (144 channels). Each raft's FEE is independently powered and contained within a conductive copper housing, and rafts are clocked with timing synchronous to < 10ns. It is expected that CCD-to-CCD crosstalk will be less than intra-CCD crosstalk, and that inter-raft crosstalk will be negligible.

**3.2 Measurements**

In a system with $N$ channels crosstalk is characterized by an $N \times N$ matrix **X** where the element $X_{ij}$ denotes the fraction of the signal in aggressor channel $i$ seen in victim channel $j$. Diagonal elements of **X** are unity and the matrix is not necessarily symmetric. Strongest couplings are expected to be between near neighbour channels.

**3.2.1 Simultaneous multi-aggressor method**

Crosstalk is usually measured in the laboratory by illuminating one segment of the CCD with an artificial star and searching for signals in the victim channels. For an $N$-channel sensor this has the drawback that $N$ images must be acquired to obtain the full crosstalk matrix. Instead, we imaged a photomask which projects 16 artificial stars, one in each segment, whose positions are staggered to keep prevent ghosts from appearing close to aggressors (Figure 5, left). The aggressor spots produced by our imaging lens are about 50 pixels in radius with about a 5-pixel wide transition region to the dark background. By using this multi-aggressor mask, the complete crosstalk matrix and nonlinearity can be measured with a single image. The efficiency of this method allows us to acquire multiple images and coadd them for increased sensitivity. An example of a 100-image stack is also shown in Figure 5, where the background noise level is about 1.2e$^-$ and crosstalk of a few parts in $10^5$ can be measured.

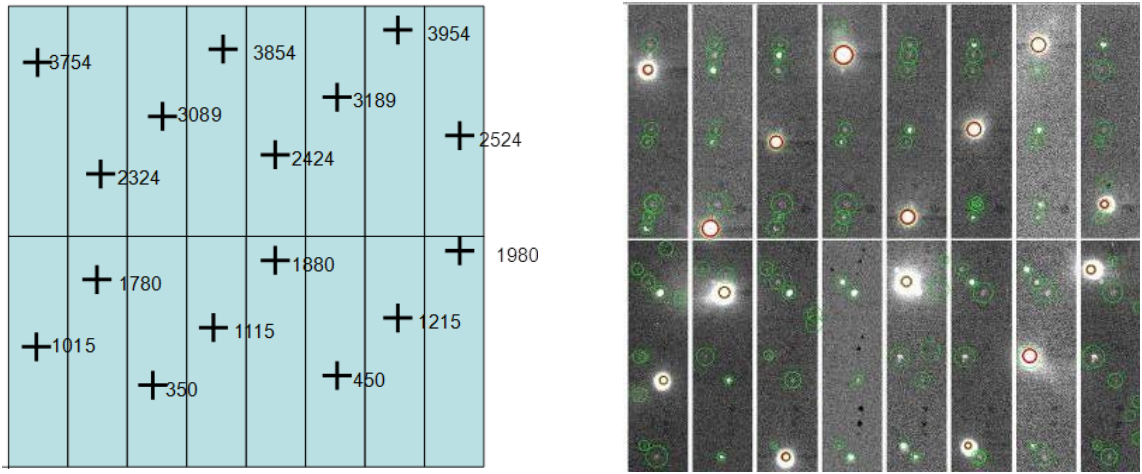

**Figure 5: Left, location of apertures in the MA mask. Right, 100-image coadd with aggressors and positive ghosts marked by red and green circles, respectively. Mask magnification factor is slightly greater than 1-to-1 and the aggressor mask has a blocked opening in one segment (bottom row, 4$^{th}$ from left).**

To measure crosstalk linearity, we construct a pixel-by-pixel scatterplot of the aggressor and victim values. An example is shown in Figure 6. Pixel values in the aggressor footprint can cover a range from near full well to zero. In this data set, nonlinearity in aggressor-to-victim coupling is not seen up to about one-fifth of full well.

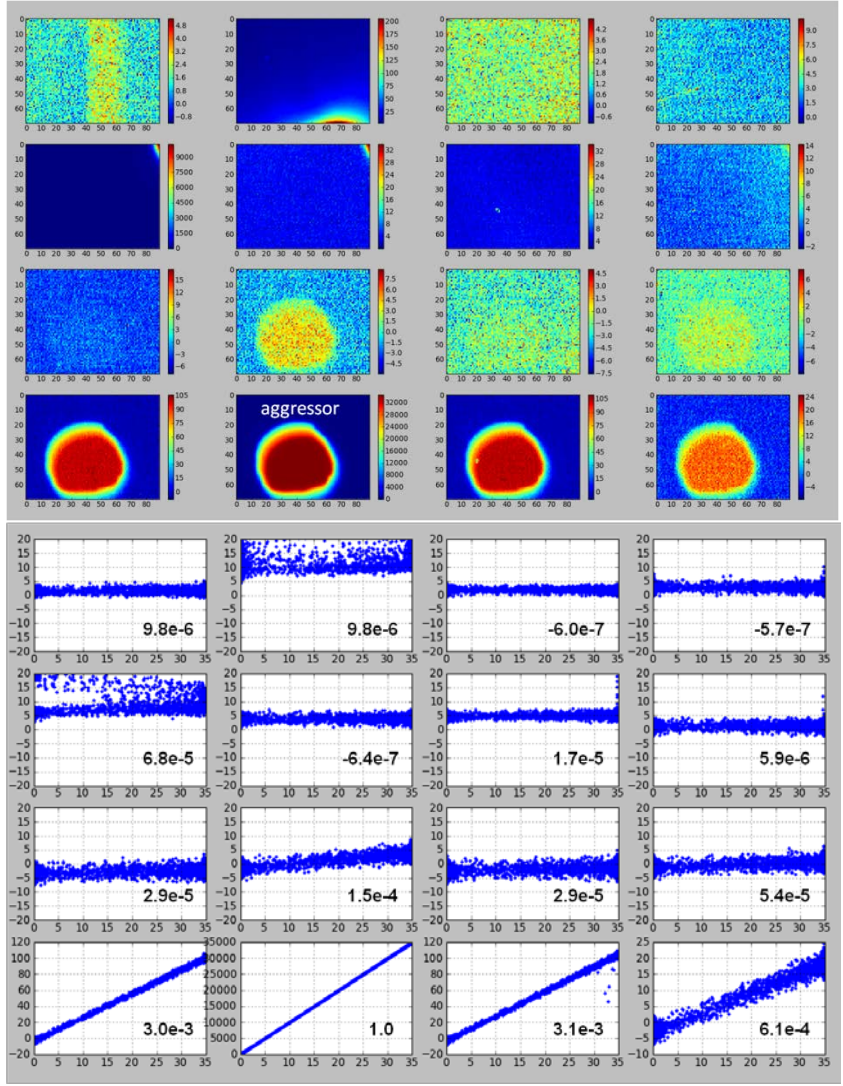

**Figure 6:** Top, zoomed regions of the MA mask image in each segment. Aggressor is in segment 14. (A portion of the segment 2 aggressor is seen in the analysis box). Bottom: pixel-by-pixel scatterplot of pixel values in the victim segments versus corresponding pixels in the aggressor segment. X-axis in kADU for segment 14. Linear fit slope displays the crosstalk amplitude.

### 3.2.2 Results

We measured the crosstalk matrix for single LSST prototype sensors in two configurations: with and without electronics located near the CCD in the cryostat. The results are shown in Figure 7. On the left the crosstalk reaches a maximum value of around $4 \times 10^{-3}$, and a pronounced 4x4 pattern is seen. The main contributor in this configuration is trace-to-trace capacitive coupling in the long (~330mm) cable connecting the CCD to the external electronics through the vacuum feedthrough. When the electronics is placed inside the cryostat with the CCDs, the cable length is only ~65mm and the crosstalk goes down by a factor of 4-5 to an acceptable level of $8 \times 10^{-4}$. SPICE circuit simulations using careful measurements of inter-trace capacitance and a vendor-supplied model of the CCD output amplifiers verify that these crosstalk levels and patterns are expected.

– 5 –

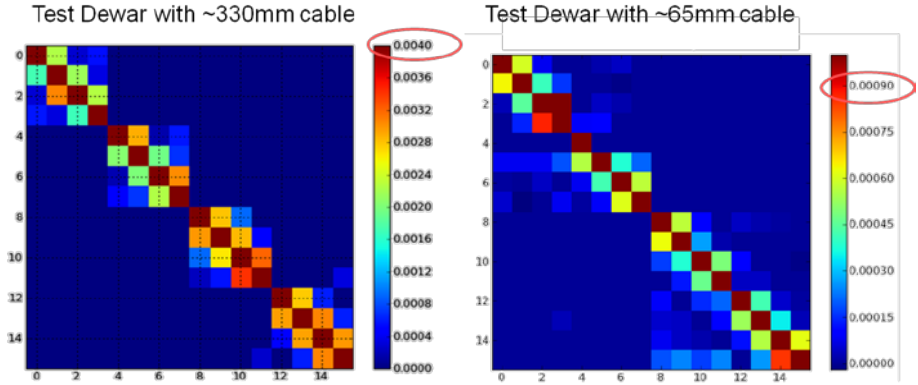

**Figure 7: Crosstalk matrix for single LSST CCD in two configurations.**

### 3.2.3 Mitigations

We modified the CCD timing sequence to insert an adjustable delay $T_d$ between charge transfer to the sense node and the start of the second DSI integration. The results are shown in Figure 8. As expected from the electrical circuit model, the crosstalk falls exponentially with $T_d$ (see Figure 3). The lower scale of Figure 8 shows the increase in frame readout time that would result from inserting delay in the pixel read sequence.

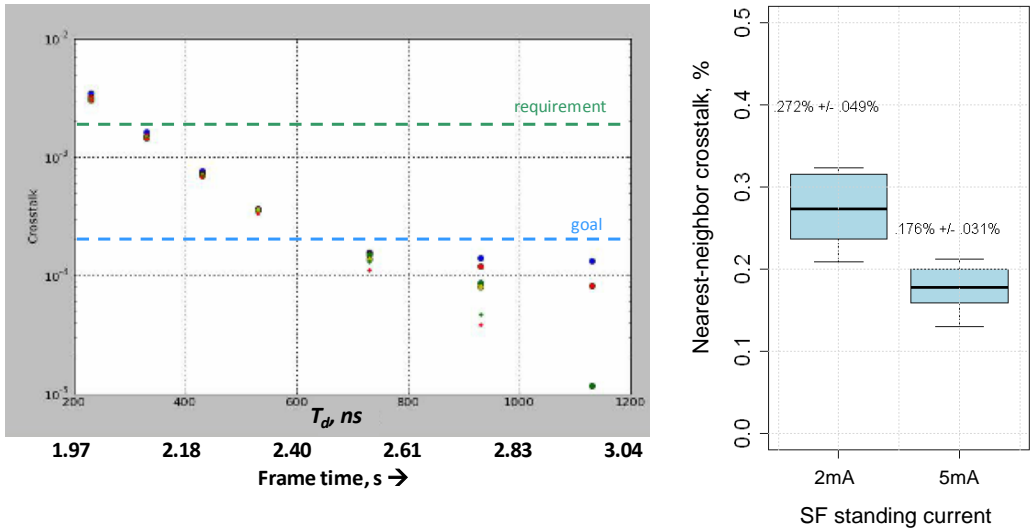

**Figure 8: Left, crosstalk amplitude (worst-case nearest-neighbor) versus isolation delay. Right, crosstalk for two levels of standing current in the output source followers. In each case the largest elements of the crosstalk matrix are shown.**

A second test was to vary the quiescent current in the output source follower amplifiers on the CCD. With an increase in standing current from 2 to 5 mA the worst-case crosstalk decreased by about 35%. This is again in agreement with the electrical circuit model: higher current increases the MOSFET transconductance resulting in a lower output impedance of the drivers, reducing the transition time of the aggressor and stiffening the output of the victim amplifier.

# 4. Implications for survey cadence

We have shown that electronic crosstalk in our multi-segment CCDs is primarily due to coupling between nearby amplifiers. Image artifacts (ghosts) produced by this mechanism will always be located at pixel coordinates which correspond to the location of the aggressor source. That is, the ghosts produced by an aggressor source at coordinate ($x,y$) in segment $i$ will occur at the same location ($x,y$) in segments $i-1$ and $i+1$, where $y$ is the number of vertical shifts and $x$ is the number of horizontal shifts between the source position and the output amplifier. Hence the ghost-to-aggressor vector in sky coordinates will depend on the readout direction of the horizontal (serial) registers in neighboring sections. An example is given below.

In Figure 9 we show the readout geometries of CCDs used in the Dark Energy Camera, HyperSuprime camera, and LSST. Note that in the two precursor surveys neighboring segments have serial registers which read out in *anti-parallel* directions, while in LSST neighboring segments read out in the *same* direction.

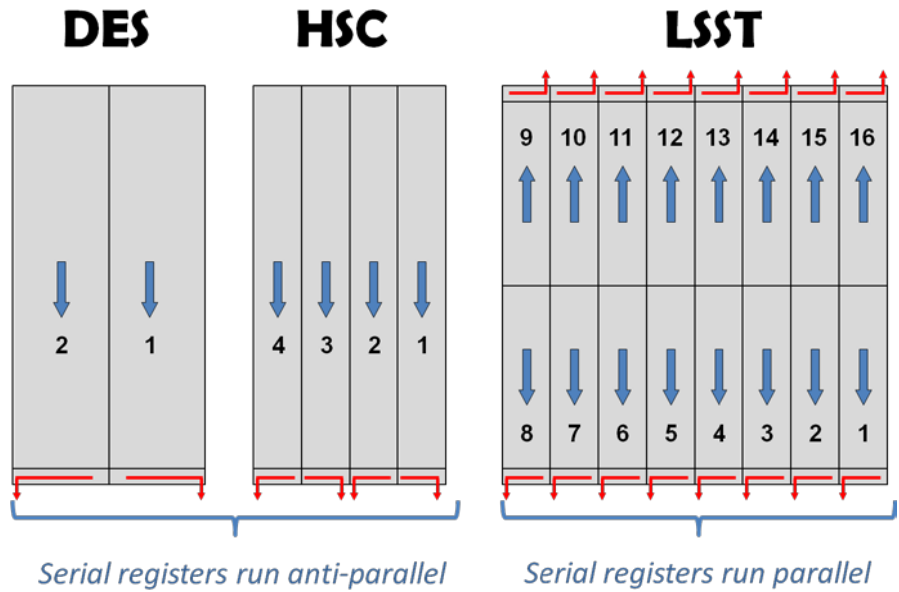

**Figure 9: Readout geometry of 2K x 4K DES and HSC sensors and 4Kx4K LSST sensor.**

Figure 10 illustrates the change in ghost-to-aggressor position when the telescope is repointed. For simplicity we show only a pair of neighboring segments of each sensor type. Consider a field with a strong source in the left-hand segment, imaged in different pointings. When the two exposures are registered and co-added, the ghosts appear separated in the DES/HSC case (a), but they are superimposed in the LSST case (b). With repeated dithers in case (a) the locus of ghost positions will be a line starting at the aggressor and extending to the right as far as the maximum x-dither or the width of the segment, whichever is greater.

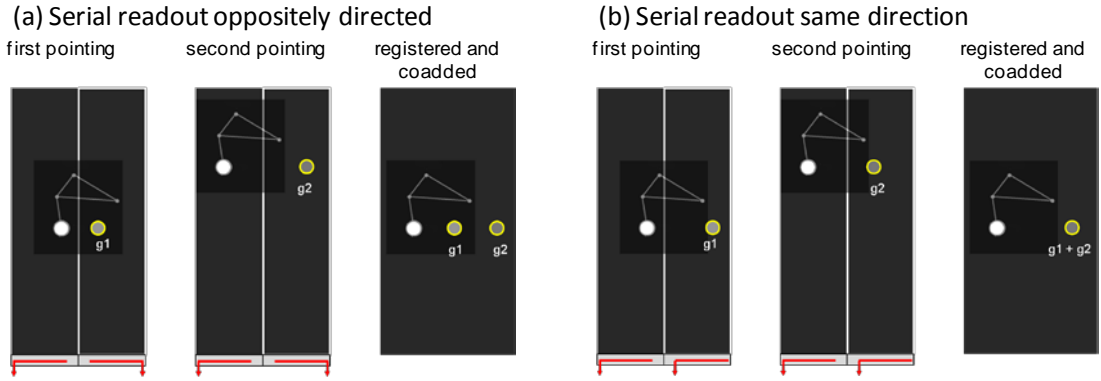

**Figure 10: Effect of crosstalk when coadding dithered images. In (a) the serial registers have antiparallel readout directions and the ghost-to-aggressor vector changes with dithering. In (b), the LSST case, ghosts track the aggressor translation and are superimposed in the coadd.**

The previous discussion pertains to dithers which preserve the rotation angle of the camera relative to sky coordinates. Dithers which include rotation will distribute the crosstalk ghosts as shown in Figure 11. Here, the locus of ghost positions for the LSST case will be a circle centered on the aggressor.

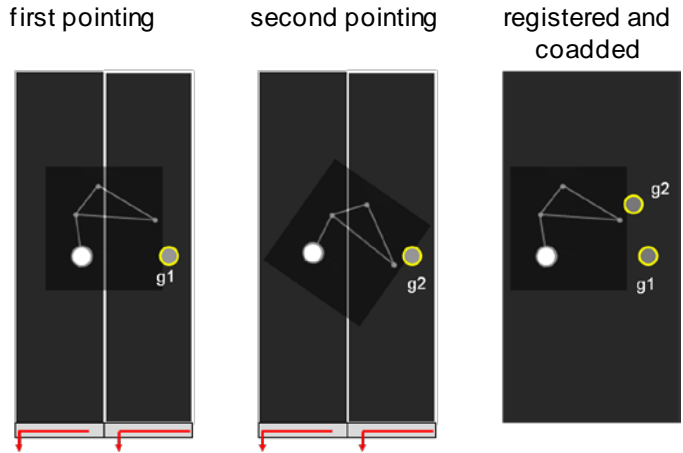

**Figure 11: Here the readout geometry of Figure 9(c) is considered for dithers that include rotation. Including rotation distributes the crosstalk ghosts that would otherwise be superimposed in the registered and coadded image.**

## 5. Summary

Electronic crosstalk in LSST will be non-negligible due to power, speed, and space constraints. An efficient measurement method using multi-source mask allows full 16x16 matrix and nonlinearity to be simultaneously acquired. Crosstalk mitigation is available, at the cost of lower frame rate or higher power dissipation. Unlike in precursor surveys, the sensor geometry in LSST will cause crosstalk to add coherently unless camera rotation with respect to sky coordinates changes with each revisit.

**Acknowledgments**

Andrei Nomerotski made measurements of capacitances in the LSST sensor prototypes, Rolf Beuttenmuller fabricated the multi-aggressor mask, and Justine Haupt set up the optical arrangement for crosstalk measurement.

This manuscript has been co-authored by employees of Brookhaven Science Associates, LLC., Portions of this work are supported by the Department of Energy under contract DE-SC0012704 with Brookhaven National Laboratory. LSST project activities are supported in part by the National Science Foundation through Governing Cooperative Agreement 0809409 managed by the Association of Universities for Research in Astronomy (AURA), and the Department of Energy under contract DE-AC02-76-SFO0515 with the SLAC National Accelerator Laboratory. Additional LSST funding comes from private donations, grants to universities, and in-kind support from LSSTC Institutional Members.